\documentclass[journal]{vgtc}                     %

\pdfoutput=1

\onlineid{1831}

\vgtccategory{Research}

\vgtcpapertype{Systems \& Rendering}

\title{HiRegEx: Interactive Visual Query and Exploration of Multivariate Hierarchical Data}

\author{%
  \authororcid{Guozheng Li}{0000-0001-6663-6712},
  \authororcid{Haotian Mi}{0009-0004-0466-4709},  
  \authororcid{Chi Harold Liu}{0000-0002-0252-329X},
  \authororcid{Takayuki Itoh}{0000-0002-1997-4644}, and
  \authororcid{Guoren Wang}{0000-0002-0181-8379}
}

\authorfooter{
  \item
  	Guozheng Li, Haotian Mi, Chi Harold Liu, and Guoren Wang are with Beijing Institute of Technology. Chi Harold Liu is the corresponding author. 
  	E-mail: \{guozheng.li, haotian.mi, chiliu, wanggr\}@bit.edu.cn
  \item
  	Takayuki Itoh is with Ochanomizu University. 
  	E-mail: itot@is.ocha.ac.jp.
}

\abstract{%
When using exploratory visual analysis to examine multivariate hierarchical data, users often need to query data to narrow down the scope of analysis. However, formulating effective query expressions remains a challenge for multivariate hierarchical data, particularly when datasets become very large. 
To address this issue, we develop a declarative grammar,  HiRegEx (Hierarchical data Regular Expression), for querying and exploring multivariate hierarchical data. 
Rooted in the extended multi-level task topology framework for tree visualizations (e-MLTT), HiRegEx delineates three query targets (node, path, and subtree) and two aspects for querying these targets (features and positions), and uses operators developed based on classical regular expressions for query construction. 
Based on the HiRegEx grammar, we develop an exploratory framework for querying and exploring multivariate hierarchical data and integrate it into the TreeQueryER prototype system. 
The exploratory framework includes three major components: top-down pattern specification, bottom-up data-driven inquiry, and context-creation data overview. We validate the expressiveness of HiRegEx with the tasks from the e-MLTT framework and showcase the utility and effectiveness of  TreeQueryER system through a case study involving expert users in the analysis of a citation tree dataset. }

\keywords{Multivariate hierarchical data, declarative grammar, visual query}

\graphicspath{{figs/}{figures/}{pictures/}{images/}{./}} %

\usepackage{tabu}                      %
\usepackage{booktabs}                  %
\usepackage{lipsum}                    %
\usepackage{mwe}                       %

\usepackage{mathptmx}                  %

\usepackage{cite}
\usepackage{amsmath,amssymb,amsfonts}
\usepackage{graphicx}
\usepackage{textcomp}

\usepackage{algorithm}
\usepackage{algpseudocode}
\usepackage{subfigure}
\usepackage{gensymb} 
\usepackage{amsfonts}

\newcommand{\oldtext}[1]{}

\begin{document}
\maketitle

\section{Introduction}
\label{sec:introduction}
Multivariate hierarchical data are ubiquitous in real-world applications and can be found in datasets like citation trees of publications~\cite{Litvis2023Tian}, reposting trees in social media~\cite{diefenbach2011formal, 2021-HierarchicalMultivariateData, 2014-WeiboEvents-Ren}, and hierarchical tabular data~\cite{li2024insightable, li2023hitailor, li2024CoInsight}. 
One technique often used to analyze such hierarchical data is exploratory visual analysis (EVA), which involves examining data, extracting patterns, gaining insights, and refining hypotheses~\cite{battle2019characterizing}.
Visual analytics techniques, such as visual encoding and querying, can facilitate the EVA process by enabling rapid specification of data visualizations and transformations~\cite{battle2019characterizing, 2022-Tarique-cacm}.
While significant progress has been made in the visual encoding of hierarchical data visualizations~\cite{schulz2011treevis}, visual querying remains a challenge in the EVA of multivariate hierarchical datasets. Specifically, the unpredictable characteristic of an EVA process indicates that users often lack a clear idea of query targets and must continuously try different queries to reach a goal.
However, the complexity of multivariate hierarchical data, in terms of topological structures and node attributes, makes constructing practical query expressions time-consuming and error-prone.

Take an example of analyzing a citation tree dataset with EVA. Each node in the citation tree represents a publication with multiple attributes, such as ``topics'' and ``authors'', while the links among nodes signify their reference relationships. In this scenario, the tasks of a researcher include capturing important publications, identifying their various patterns, and comprehending the development of research trends. 
The researcher needs to formulate diverse query expressions frequently to accomplish tasks or validate hypotheses.
For instance, the researcher may query publications within the ``graph'' topic from the past three years that have been cited by more than five publications within the ``immersive'' research topic to understand the recent intersection of disciplinary directions. 
The query expression is related to both node attributes, like topics, and topological structures, such as the number of children on specific topics.
In particular, the topics and parameters in the above query might be changed frequently during EVA.

Existing techniques allow users to query multivariate hierarchical data by programming and interactive filtering.
The programming approach lets users craft general low-level imperative codes, which can be easily verified but 
requires clearly defined goals and query targets, a significant burden for non-programming users. 
Although some declarative languages for hierarchical data queries~\cite{levy2006tregex, beyer2011jaql, slingsby2009configuring} are less challenging, they were usually domain- or task-specific (\textit{e.g.}, the syntax tree in natural language processing~\cite{levy2006tregex}, or graph query languages~\cite{rodriguez2015gremlin,francis2018cypher, van2016pgql,angles2018g}) and do not adequately support diverse visual analytics tasks~\cite{2021-treetask-Pandey}. 
With the interactive filtering approach, on the other hand, users can see specific topology structures and build dynamic queries on multiple node attributes. However, constructing flexible and diverse queries requires continuous data filtering and checking activities, a very time-consuming task, especially when the dataset is large.
Therefore, we seek an efficient and expressive query approach, which can be integrated into the EVA process for multivariate hierarchical data.

This work presents HiRegEx (\textbf{Hi}erarchical data \textbf{Reg}ular \textbf{Ex}pression), a novel declarative grammar for querying multivariate hierarchical data.
HiRegEx builds upon a tree-specific extension to the multi-level task topology framework (e-MLTT)~\cite{2021-treetask-Pandey}, a dataset with a collection of 213 tasks from existing studies. 
We develop a classification of three distinct targets (node, path, and subtree) from the task abstraction framework and then define two aspects of target querying, features, and positions.
Specifically, HiRegEx sees a node as the elementary unit, on which users can set constraints on various attributes and borrow operators from regular expressions to specify parent-child relations~\cite{schulz2010design}. Some new operators are also introduced to specify sibling relations~\cite{li2020gotree}.

Furthermore, we introduce a query-based framework to support the EVA of multivariate hierarchical data based on the HiRegEx declarative grammar.
The framework is designed to assist users in conducting exploratory analysis tasks based on the sense-making mode~\cite{lee2019you, pirolli2005sensemaking}.
The framework includes top-down pattern specification, bottom-up data-driven inquiry, and context-creation data overview.
We implement a prototype system, TreeQueryER, to integrate the exploratory framework.
The top-down specification is supported by a visual editor based on HiRegEx, in which users can construct query expressions interactively. 
To conduct bottom-up inquiries, users can test different query expressions based on the visualized target dataset. 
The context-creation overview shows the hierarchical data collection through the dimension reduction technique and incorporates a graph contrastive learning model for the computation of similarities among distinct data.

We validate our techniques in two ways.
First, we demonstrate the expressiveness of HiRegEx grammar based on the e-MLTT framework for hierarchical data. 
The results indicate that the HiRegEx grammar effectively supports 174 tasks of the entire task collection, consisting of 213 tasks.
Second, we validate the utility of the TreeQueryER prototype system in a case study involving a citation tree dataset. 
The results show that users can interactively construct query expressions for various tasks and confirm the capacity of the underlying framework to facilitate the EVA process of multivariate hierarchical data.

In summary, the main contributions of this paper are as follows: 
\begin{itemize}[leftmargin=3.5mm]
\item The HiRegEx declarative grammar to facilitate diverse tasks in querying multivariate hierarchical data; 
\item A query-based exploratory framework for exploring multivariate hierarchical data, supporting top-down, bottom-up, and context-creation data query modes; 
\item The TreeQueryER prototype system integrates the exploratory framework for multivariate hierarchical data and is validated through a case study involving a citation tree dataset. Our source code is available at \url{https://github.com/bitvis2021/HiRegEx}.
\end{itemize}

\section{Related Work}
In this section, we review literature in the areas of query languages of hierarchical data and interactive visual query techniques.

\subsection{Query Languages of Hierarchical Data}

Hierarchical data query has been extensively studied due to its wide applications in various domains.
A typical application scenario is for querying syntax trees, which are hierarchical representations of different syntactic categories of a sentence, in natural language processing.
Tregex~\cite{levy2006tregex} is a query language for syntax trees based on character matching and supports the specification of relative relationships between nodes. 
However, dealing with a large-scale syntax tree with this approach involves continuously integrative processes, which can be computationally complex and produce results with poor readability.
In addition, unlike the traditional usage scenarios of hierarchical data queries, the design of Tregex emphasizes the order between sibling nodes, imposing some challenges for many users. 
Moreover, many databases contain data arranged in a hierarchical structure.
Jaql~\cite{beyer2011jaql} is a declarative scripting language most commonly used for querying and processing JSON data, a classic format used for hierarchical data.
The hierarchical query language (HQL)~\cite{slingsby2009configuring} is a language for querying a hierarchical database. 
Based on HQL, users can specify transactions against a database with a hierarchical structure. 
Both Jaql and HQL are designed for users to set the attribute value for querying targets, such as single node attributes or aggregated node attributes. However, these techniques do not support the specification of the topological structure.

Given a tree as a connected acyclic undirected graph, the querying language for general graphs can be used in most situations.
Cruz et al.~\cite{cruz1987graphical} proposed a declarative language called G to support regular expressions in specifying the path between any two nodes in a graph.
G+~\cite{curz1988g} extends G with a summary graph to restructure the results obtained by a query statement. 
Furthermore, GraphLog~\cite{consens1990graphlog} is a query language for hypertext based on G+, and adds negation and unifies the concept of a query graph. 
These graph query languages~\cite{cruz1987graphical, curz1988g, consens1990graphlog} use regular expressions to define the path between nodes and allow the concise specification of relationships between two non-adjacent nodes within a large-scale graph. 
In addition to the languages defined over a simple graph model, industry users also adopt other commonly used graph database query grammars, such as Gremlin~\cite{rodriguez2015gremlin}, Cypher~\cite{francis2018cypher}, PGQL~\cite{van2016pgql}, and G-CORE~\cite{angles2018g}. 
These languages have good expressiveness and support the specification of topological structures and node attributes.
However, existing graph query languages cannot specify the sibling or same-level relations expressively, which is an essential feature of hierarchical data. 
In addition, the aforementioned imperative query languages are not intuitive for ordinary users to complete query tasks for multivariate hierarchical data.

Despite the availability of many query languages for filtering hierarchies across various scenarios, they often lack the flexibility to support users in querying multivariate hierarchical data for EVA.

\subsection{Interactive Visual Query Approach}
In visualization research, efforts on interactive visual queries involve nearly all data types, such as tabular data, time-series data, event sequences, and graphs. 
The following discussions emphasized the closely relevant data types with hierarchies:  graph, time series, and event sequence.  
More specifically, hierarchy is a special graph data with hierarchical relationships and no circles. 
In addition, both time series and event sequences have a linear structure, similar to the decomposition paths of hierarchies, linked from the root node to the leaf nodes.

\textbf{Graph data query}. 
For the visual query of graph data, Graphite~\cite{chau2008graphite}, VOGUE~\cite{bhowmick2013vogue}, and Visage~\cite{pienta2016visage} allow users to construct query patterns for graph data intuitively, and these methods present the matching subgraphs of a large property graph. 
Vertigo~\cite{2022-VERTIGo-Cuenca} allows users to build and recommend graph queries and delve into their findings within multi-layer networks. 
In contrast, VIGOR~\cite{2018-VIGOR-Pienta} emphasizes the efficient summarization of subgraph queries by grouping results according to node characteristics and structural similarity.
VIMO~\cite{troidl2023vimo} improves the intuitiveness and flexibility of graph queries by allowing users to sketch the targets in the query interface and enable users to define structural similarity constraints directly.
Furthermore, to reduce the difficulty in querying graphs, researchers have paid more attention to the data-driven graph query approaches and developed methods like selecting high-quality patterns (small subgraphs) automatically from the underlying graph database~\cite{bhowmick2020aurora, huang2021midas, yuan2021towards}. 
All these methods focus on how to specify the graph topology accurately by constructing subgraph patterns. However, different from graph data, hierarchical data consists of both parent-child and sibling relations. Because of the lack of the capability to distinguish these two types of relations, existing graph query methods fall short on tasks of multivariate hierarchical data.

\textbf{Time series query.} The most straightforward approach for the visual query of time series is to interactively select a partial time series as the query. 
Users can brush the partial time series of interest through timeboxes~\cite{hochheiser2003interactive, hochheiser2004dynamic, buono2005interactive, buono2008interactive} on a two-dimensional space. 
The extent of the timebox on the time axis (\textit{x}-axis) specifies the time of interest, while the extent on the value axis (\textit{y}-axis) specifies a constraint on the range of values of interest within the time. 
These methods allow users to draw multiple timeboxes interactively to query a significant amount of time series data simultaneously.
Querying time series using timeboxes requires users to understand the data clearly.
To solve this problem, QuerySketch~\cite{wattenberg2001sketching}, QueryLines~\cite{ryall2005querylines}, and Qetch~\cite{mannino2018qetch} query time series data using sketching. 
Users can directly draw on the visualized time series to find similar instances, and Qetch~\cite{mannino2018qetch} combines freehand drawing and regular expressions to query time series data. 
In addition to sketching patterns, COQUITO~\cite{krause2015supporting} provides a visual interface that allows users to define cohorts with temporal constraints through intuitive drag-and-drop operations, enabling the exploration and analysis of time series data with specific temporal patterns.

The data-driven approach plays an essential role when users cannot precisely articulate the targets. 
Peax~\cite{lekschas2020peax} is the first deep-learning-based approach for an interactive visual pattern search in sequential data.
It uses a convolutional autoencoder to capture more visual details of complex patterns.
Zenvisage++~\cite{lee2019you}, which finds that sketch-based methods are not always efficient, develops a taxonomy of visual query system capabilities,  including a top-down pattern search (translating a pattern ``in-the-head'' into a visual query), bottom-up data-driven inquiries (querying or recommending based on data), and context-creation (navigating across different collections of visualizations). Our effort to develop an exploratory framework for multivariate hierarchical data is partially inspired by this taxonomy.

\textbf{Event sequence query.} For the interactive visual query of event sequence data, (s$|$qu)eries~\cite{zgraggen2015eries} is a multi-attribute event sequence data query method based on regular expressions. 
Users can interactively construct their query pattern by adding various constraints on a node-link diagram. 
EventPad~\cite{cappers2018eventpad} also uses regular expressions to query event sequence data. 
The difference between these two techniques is that EventPad defines query results as abstract behavior rather than for data extraction, and supports the aggregation, alignment, and combination of event sequence data. 
VESPa 2.0~\cite{krueger2017vespa} facilitates the discovery and validation of movement sequence patterns through interactive exploration and querying, supporting a bottom-up and data-driven exploration approach. 
Beyond the query task, MAQUI~\cite{law2018maqui} offers a novel approach for recursive event sequence exploration, combining querying and pattern mining to help users analyze and understand complex event sequence data more effectively.
Similar to event sequence data, orders exist between data items in the hierarchical data because of the parent-child relationships. 
Therefore, hierarchical data can be decomposed into a collection of multiple sequences, and the specification of hierarchical data can borrow ideas from regular expressions. 

Visual queries play a crucial role in enabling EVA and offer flexible filtering capabilities. While visual query techniques for data types with relational and linear structures have been extensively studied, there remains a gap in addressing visual query for hierarchical data.

\section{Hierarchical Data Query Task Space}
\label{sec:task-space}

In this section, we define the task space related to querying multivariate hierarchical data. Pandey \textit{et al.} comprehensively summarized analytical tasks for tree visualizations and introduced the e-MLTT framework~\cite{2021-treetask-Pandey}, which enhances the specificity of task abstraction definitions tailored to tree visualizations. The e-MLTT framework decomposes tree visualization tasks along two dimensions: targets and actions (see Fig. 4 in Pandey's study~\cite{2021-treetask-Pandey}). The target is defined as the object related to the tree visualization tasks, while the action is the operation users performed to accomplish the tasks. Our task space definition is also based on these two dimensions of the e-MLTT framework, but only keeps the query-related analytical tasks in the framework. In this section, we detail the targets and actions in our task space, using the citation tree as an example to explain relevant concepts and techniques.

\subsection{Query Targets}
\label{sec:query-targets}
A query target is defined as the object that the query expression aims to match. 
The traditional multi-level task typology (MLTT) framework proposed by Brehmer and Munzner~\cite{brehmer2013multi} categorizes the target of tasks into ``\textit{topology}'' and ``\textit{attributes}''.
The e-MLTT framework~\cite{2021-treetask-Pandey} further extends MLTT to include more specificity to support tree-specific tasks. 
It divides the specific targets of ``topology'' for tree visualization tasks into four categories: tree, subtree, path, and node, and the ``attributes'' into two categories: \textit{node attributes} and \textit{link attributes}.
The query targets of the HiRegEx grammar fully consider the targets defined in the e-MLTT framework, except for link attributes. This is because links in hierarchical data are typically used to represent parent-child relationships, and the e-MLTT framework~\cite{2021-treetask-Pandey} also indicates that links are less frequently used for encoding data attributes in tree visualizations, and tasks related to links are rare in tree visualizations.

\begin{figure}[tb]
 \centering 
 \includegraphics[width=\columnwidth]{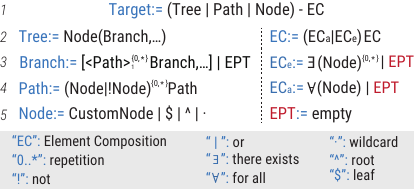}
 \caption{The formal specification of the HiRegEx declarative grammar. The first row introduces the overall structure of the HiRegEx specification. Rows 2 to 5 below present various query targets on the left, grounded in the e-MLTT framework, while the right side details element compositions that enable users to specify how these query targets are structured.}
 \label{fig:formal-specification}
\end{figure}

\subsection{Query Actions}
\label{sec:query-actions}
A query action is an operation users perform to accomplish a task.
The e-MLTT framework classifies query actions into three levels (high, mid, and low). 
Our focus here is on the mid-level actions---``\textit{search}'', a type of operations a user must perform to find targets. 
Under e-MLTT, the ``search'' action is further divided into four subtypes ---``\textit{lookup}'', ``\textit{locate}'', ``\textit{browse}'', and ``\textit{explore}''---based on whether the knowledge of targets and locations is available. 
In particular, search in the ``\textit{target known}'' subtype refers to tasks with explicit knowledge of a target's identity. 
For example, when users want to find some popular papers in certain directions that have attracted much attention, they can define a query expression to find publications with more than five iterative citations within a short period.
Conversely, the ``\textit{location known}'' search concerns tasks that clearly describe a target's position within the tree.
For instance, to understand the ongoing development of a technique, users can look for the inaugural paper that introduced the technique and then trace a subsequent path in the citation tree. As a result, all papers along this path are likely to be related to this technique.

\section{Query Grammar Design}
\label{sec:query-grammar-design}
In this section, we first analyze the design requirements for query grammar based on the task space for querying multivariate hierarchical data and then present the detailed grammar specifications of HiRegEx.

\subsection{Design Requirements}
\label{sec: Design Requirements}
Existing studies~\cite{2006-visql-Hanrahan, 2016-vegalite-satyanarayan, 2018-atom-tvcg, kim2020gemini} have significantly influenced and shaped the following two requirements for a declarative grammar designed for querying multivariate hierarchical data.

\textbf{R1: Expressiveness}. A visualization grammar should have the capability to articulate the entire design space~\cite{2018-atom-tvcg, fowler2014dsl}. 
Hierarchical data is a generic data type with broad applications in areas such as finance, biology, and computer science~\cite{li2020tvcg, li2023gotreescape}.
This implies that the grammar should support versatile data query functions applicable to various scenarios involving hierarchical data.
The e-MLTT framework~\cite{2021-treetask-Pandey} has summarized 213 analysis tasks related to the hierarchical data from the literature.
As mentioned in Sec.~\ref{sec:task-space}, query targets in the task space consist of topology and attribute, while query actions are divided into four types according to whether ``target'' and ``location'' are known. 
Consequently, the declarative query grammar should empower users to obtain different kinds of query targets based on various query actions, aligning with the requirements of most tree visualization tasks defined by the task abstraction model.

\textbf{R2: Conciseness}. 
EVA is a typical usage scenario of query languages for multivariate hierarchical data. In the EVA process, analysts often begin with an unclear goal and refine their objectives as they explore the data~\cite{wongsuphasawat2015voyager}. This process necessitates frequent data queries for effective exploration~\cite{gotz2008characterizing, siddiqui2016effortless}.  To improve the efficiency of EVA, analysts prefer query expressions that can be easily constructed, such as higher-level grammar like ggplot2~\cite{wickham2010layered} and grammar-based systems like Tableau~\cite{Stolte2002Polaris}. 
In addition, analysts need to understand the previous exploratory analysis process to make informed decisions about further exploration. Therefore, it is crucial to ensure the conciseness of the grammar, making it easy for users to understand and construct. The query expressions should align with users' cognitive understanding of analysis tasks to enhance readability and construction efficiency.

\subsection{Grammar Specification}
\label{sec:grammar-specification}

Existing research has extensively explored semi-structured data querying~\cite{levy2006tregex, beyer2011jaql, slingsby2009configuring,rodriguez2015gremlin,francis2018cypher,angles2018g}. However, these techniques either lack the expressive power to cover all tree analysis tasks or produce non-intuitive expressions (\textbf{R1}, \textbf{R2}), making them unsuitable for EVA scenarios.
Since a tree can be viewed as a composition of multiple paths, and regular expressions provide an intuitive way to represent paths, we extend regular expressions to support various tasks in multivariate hierarchical data querying~(\textbf{R1}). We present HiRegEx, a declarative grammar for querying multivariate hierarchical data. Regular expression uses ordinary characters (``a'' to ``z'') and special operators to define text sequences conforming to a pattern, commonly employed for text querying and replacement. Similarly, HiRegEx is designed to search multivariate hierarchical data by specifying patterns through nodes with attribute constraints and special operators. The operators in HiRegEx are borrowed from the regular expression and further extended according to the characteristics of the hierarchical data. 
The simplicity of operators ensures the conciseness of the query expression~(\textbf{R2}).

The specifications of HiRegEx grammar (see Fig.~\ref{fig:formal-specification}) support all query targets in the e-MLTT framework, node, path, subtree, and tree, as explained in Sec.~\ref{sec:query-targets}~(\textbf{R1}).
Additionally, we also introduce element compositions to allow users to specify the compositions of various query targets.
Users need to specify different query patterns for various targets.
For node queries, we introduce various constraints for node attributes.
Querying paths necessitates the definition of parent-child relationships based on node specifications.
Moreover, querying subtrees and trees entails specifying the complete topological structure, including parent-child and sibling relationships.

In addition to query targets, our query grammar is also designed to support different query actions defined in the task space~(\textbf{R1}).
For ``\textit{target known}'' queries, the grammar assists users in specifying the inherent features of the target, encompassing topology structures, node attributes, and element compositions. 
Furthermore, to facilitate ``\textit{location known}'' queries, the grammar enables users to precisely define the target's location within the entire hierarchical dataset, thereby representing its positional features.

\textbf{(1) Node Query.}
\label{sec:node-query}
A node is the elementary unit of hierarchical data and has multiple attributes.
The node expression within HiRegEx enables users to specify patterns for node attributes.
We divide the node attributes into two categories: inherent and additional attributes.
\begin{itemize}[leftmargin=3.5mm]
\item \textbf{Inherent attributes}, often quantitative data like depth, help users in specifying target positions (location-known query). 
In addition, each node can also be seen as the root of a subtree. Therefore, HiRegEx allows users to specify the tree-specific attributes in the node, including size, height, and width. 
At the same time, the parent-child relationship defines a strict order between nodes, and each node is in a unique path starting from the root node. 
HiRegEx facilitates users in specifying node attributes according to their related nodes, with operators ``$\&$''  and ``$\#$'' for relative and absolute positions across levels, respectively. For instance, ($\textit{degree}=\&$-1) signifies that the degree of a node is the same as that of its parent, and ($\textit{degree}=\#1$) implies equivalence to the degree of the root node.

\item \textbf{Additional attributes} are often quantitative and categorical node features, such as citation number and authors in the citation tree (target-known query).
To articulate the constraints of node attributes, we introduce several predicate operators in the grammar specification, including ``$\textgreater$'', ``$\ge$'', ``$\textless$'', ``$\leq$'', ``$=$'' for quantitative data, and ``$\in$'' for categorical data.
Nodes in multivariate hierarchical data are selected if their attributes satisfy all defined constraints within the expression.
Query results of a node expression consist of individual nodes within multivariate hierarchical data.
We also pre-define three special nodes: the wildcard node ($\bullet$), root node ($\wedge$), and leaf node ($\$$). 
The following defines the formal specification of \textit{Node}:
$$Node := CustomNode\,|\,\$\,|\,\wedge\,|\,\bullet$$
\end{itemize}

\textbf{(2) Path Query.}
Based on the node specifications, we define the path-related syntax within hierarchical data by detailing the parent-child relationships. 
A path in hierarchical data exhibits a linear structure, wherein node sequence signifies parent-child relationships. 
Specifically, the preceding node serves as the parent of the subsequent nodes. 

To query paths in hierarchical data, we adopt operators from regular expressions, including ``or'' ($|$), ``not'' ($!$), and ``repetition'' (\{$min$, $max$\}).
The repetition operator empowers users to specify an exact number or a range. For instance, ${node}^2$ signifies that the $node$ pattern repeats twice, and ${node}^{\{2,5\}}$ indicates a repetition range from two to five. 
Users can specify only the minimum or maximum number of repetitions, such as ${node}^{\{2,\}}$ matching at least two times and ${node}^{\{,5\}}$ matching up to five times.
HiRegEx consistently employs a lazy matching strategy to query paths according to the expression, which means the path that first satisfies the query expression is the result, concluding the matching process.
An exception arises when the maximum number of repetitions is unspecified. In this case, the matching process concludes only if no nodes can be matched.
The formal specification of Path is presented below.
For instance, the query expression (\textit{authors}=``Ben Shneiderman'')$^{\{3,\}}$ can search for a citation path that indicates Ben Shneiderman’s continuous iterative studies in a specific research topic. 
The formal specification of $Path$ is presented below:
$$Path := (Node \;|\; !Node )^{\{min, max\}} Path$$

\begin{figure}[tb]
 \centering 
 \includegraphics[width=0.9\columnwidth]{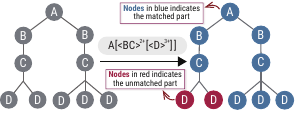}
 \caption{The explanations of \textit{Branch} operator in the HiRegEx. The nodes in blue indicate the matched part with the HiRegEx expression, while the nodes in red indicate the unmatched part.}
 \label{fig:branch-explanation2}
\end{figure}

\textbf{(3) Subtree/Tree Query.}
The above path query expressions facilitate users in specifying parent-child relationships between nodes, yet the topological structures of hierarchical data also require the determination of sibling relations~\cite{li2020gotree}. 
This section delves into the specification of sibling relationships that do not dictate the node sequence. 

The relations between nodes' siblings are implied by relations between paths. 
To specify the sibling relation, we introduce the $Branch$ operator, which merges multiple paths and allows users to specify the repetition number of paths. 
Notably, an inner path within a $Branch$ can be followed by another $Branch$, aligning with the recursive characteristics of hierarchical data. 
The formal specification of the $Branch$ operator is expressed as:
\begin{equation} 
\label{eq1}
\begin{split}
Branch := [&{Path_1}^{\{min, max\}}Branch,\cdots, \\
&{Path_n}^{\{min, max\}}Branch]
\end{split}
\end{equation}

When a path is connected with a Branch, the query results can only be determined if the matching process is finished.
An illustrative example in Fig.~\ref{fig:branch-explanation2} displays two paths (B-C) beneath node A.
However, the nodes in red do not meet the requirement. 
More specifically, the repetition number should exceed three according to the expression.
Only one path under node A satisfies the condition, falling short of the expression's demand for more than two paths.
Finally, the query expression can not be matched with the hierarchical data. 
Based on the \textit{Branch} operator, the formal specification of \textit{Subtree} is shown below. 
For example, the query expression for citation tree (\textit{authors}=``Ben Shneiderman'')[$\langle$citation$\geq$200$\rangle^{\{3,\}}$] can search for a Shneiderman's paper which has inspired more than three highly cited papers.

$$ Subtree := Node\,Branch $$

\begin{figure*}[tb]
 \centering 
 \includegraphics[width=0.96\textwidth]{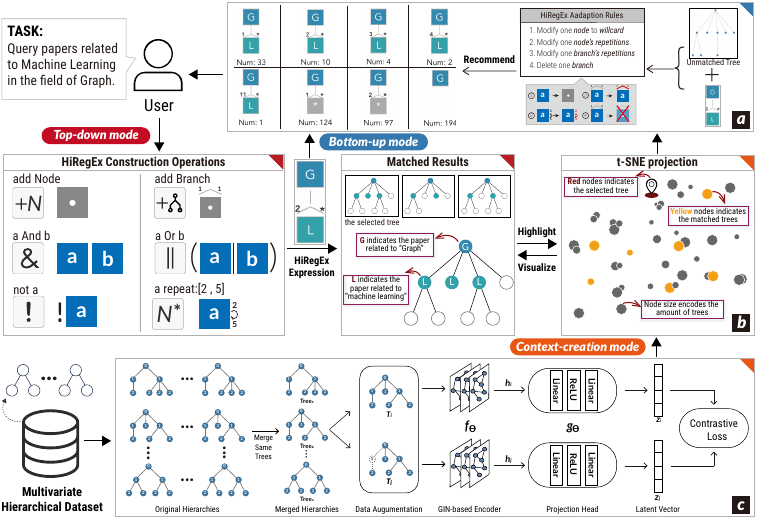}
 \caption{
 The exploratory framework for querying multivariate hierarchical data comprises three modes: \textit{top-down}, \textit{bottom-up}, and \textit{context-creation}. 
 The top-down mode starts from a clear query task. 
 Users construct the corresponding query expression through direct manipulations interactively. 
 The bottom-up mode recommends related query expressions based on the initial expression and the multivariate hierarchical data collection. 
 The context-creation mode offers users an overview of the entire hierarchical data collection. 
 Modules associated with the top-down, bottom-up, and context creation modes in the framework are denoted by red, orange, and blue triangles.
 }
 \label{fig:pipeline}
\end{figure*}

\textbf{(4) Element Composition.}  We find that specific query tasks necessitate consideration of the comprehensive element compositions of the query target (target-known query). For instance, analysts may seek to identify influential papers by querying citation trees published in 2019, comprising more than ten highly cited papers. To accommodate such query tasks, we introduce the \textbf{E}lement \textbf{C}omposition (\textit{EC}) operator in HiRegEx, empowering analysts to specify compositions as an additional aspect of the query target. The specifications of EC can be categorized into two types: for all ($\forall$, denoted as $EC_a$) and there exists ($\exists$, denoted as $EC_e$). The formal specification of \textit{EC} is presented below.
\begin{align*}
EC & := (EC_a \;|\; EC_e) \; EC\\
EC_a & := \forall \langle Path\rangle^{\{min,max\}} \;|\; EPT \\
EC_e & := \exists \langle Path\rangle^{\{min,max\}} \;|\; EPT
\end{align*}

Note that the \textit{repetition} operator(${min, max}$) in \textit{EC} refers to the occurrence number of \textit{Path}, distinguishing it from the repetition (${min, max}$) used with the \textit{Node}.
With the \textit{EC}, we can articulate the task above through the following expressions. 
\begin{align*}
(year=2019)[\langle(\bullet)^{\{0,\}}\rangle^{\{0,\}}] \text{-} \exists (citation\geq200)^{\{10,\}} 
\end{align*}

With all the previously defined operators, we can formally specify the query target, denoted as \textit{Target}, as follows.
\begin{align*}
Target := (Subtree | Path | Node) \text{-} EC 
\end{align*}

\section{Query-based Exploration Framework}
\label{sec:query-based-exploration-framework}
EVA constitutes an iterative process involving data presentation and interactive queries~\cite{battle2019characterizing}, aligning with the principles outlined in the visual information-seeking mantra~\cite{1996-Shneiderman}. This holds for multivariate hierarchical data as well. 
This section presents a query-based exploratory framework tailored for multivariate hierarchical data. 
The exploratory framework, rooted in the visual query sense-making model~\cite{lee2019you}, is designed to explore multivariate hierarchical data comprehensively. 
Illustrated in Fig.~\ref{fig:pipeline}, the framework provides analysts with three distinct modes tailored to address various requirements: top-down mode, bottom-up mode, and context-creation mode.

\begin{figure}[tb]
 \centering 
 \includegraphics[width=\columnwidth]{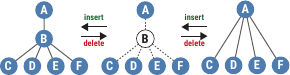}
 \caption{The \textit{delete} and \textit{insert} operations for computing tree edit distance.}
 \label{fig:edit-distance}
\end{figure}

\begin{figure}[tb]
 \centering 
 \includegraphics[width=\columnwidth]{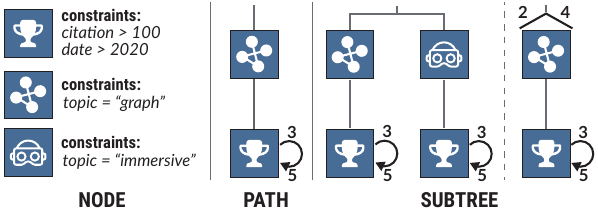}
\caption{Three visual operators in the query expression: (a) Node, (b) Path, and (c) Branch, which are the basic components of query expressions. }
 \label{fig:visual-query-operators}
\end{figure}

(1) \textbf{Top-down mode} is designed for a goal-oriented query process where users have a clear understanding of the target pattern and aim to search for data instances exhibiting the pattern.
As explained in Sec.~\ref{sec:query-grammar-design}, HiRegEx is designed based on the task space for querying multivariate hierarchical data. 
Specifically, users can convert requirements into query expressions without low-level programming.
After executing a query expression, users can obtain target data. 
The challenge in the top-down mode lies in translating the desired patterns by users into executable query expressions because constructing HiRegEx expressions in a textual format requires a steep learning curve.
More specifically, users need to memorize operators and parameters in the specification, and text-based expressions lack cognitive consistency with the query results of hierarchical data.
To address this limitation, we propose a visual operator for each component in the query expression, including the Node, Path, and Branch shown in Fig.~\ref{fig:visual-query-operators}. 

(2) \textbf{Bottom-up mode} is a data-driven process enabling users to identify something of interest from the data collection. 
In this mode, analysts are initially unclear about query targets or unable to specify query patterns accurately.
They need to determine the query expression or specific parameters according to the data collection.
The challenge in this mode lies in generating an appropriate set of stimuli through recommendations that can prompt further data-driven inquiries.

We devise an expression recommendation algorithm that can derive several relevant query patterns based on a pre-determined expression and data collection.
The recommendation algorithm checks each item in the data collection based on the initial HiRegEx expression. 
For each data item that fails to match the initial query expression, the recommendation algorithm iteratively refines the expression through the following operations until the matching process is finished (see Fig.~\ref{fig:pipeline}(a)):
(1) changing a node with several constraints to a wildcard; 
(2) modifying the repetition of one node; 
(3) modifying the repetition of one path; 
(4) deleting a path from a branch. 
These four operations are prioritized in descending order based on their impact on the matching results. 
After traversing all entries, the algorithm merges the adjusted expressions and provides them as recommendation results to users.
Details of the algorithm can be found in the supplemental material.

(3) \textbf{Context-creation mode} aims to assist users in understanding the data distribution and offering relevant data as context for each query result, thereby assisting users in subsequent exploration.
The design challenge of the context-creation mode is constructing an effective overview that reflects the similarities between hierarchical data.
The tree edit distance is a typical metric to quantify similarities in tree-structured data. It is defined as the minimum-cost sequence of node operations (\textit{e.g.}, insert) required to transform one tree into another.
However, the tree edit distance lacks consistency with the query expression matching algorithm, which traverses hierarchical data from top to bottom, as explained in Sec.~\ref{sec:grammar-specification}.
While tree edit distance might yield a small value between two hierarchical data, their topological structures can vary significantly from top to bottom due to operations that allow inserting or deleting any nodes.
For example, Fig.~\ref{fig:edit-distance} illustrates the insertion and deletion operations between two hierarchical datasets. 
The edit distance between different hierarchies is small. 
However, their topological structures differ significantly; one has only one node at the second level, while the other has four nodes.

\begin{figure*}[tb]
 \centering 
 \includegraphics[width=0.96\textwidth]{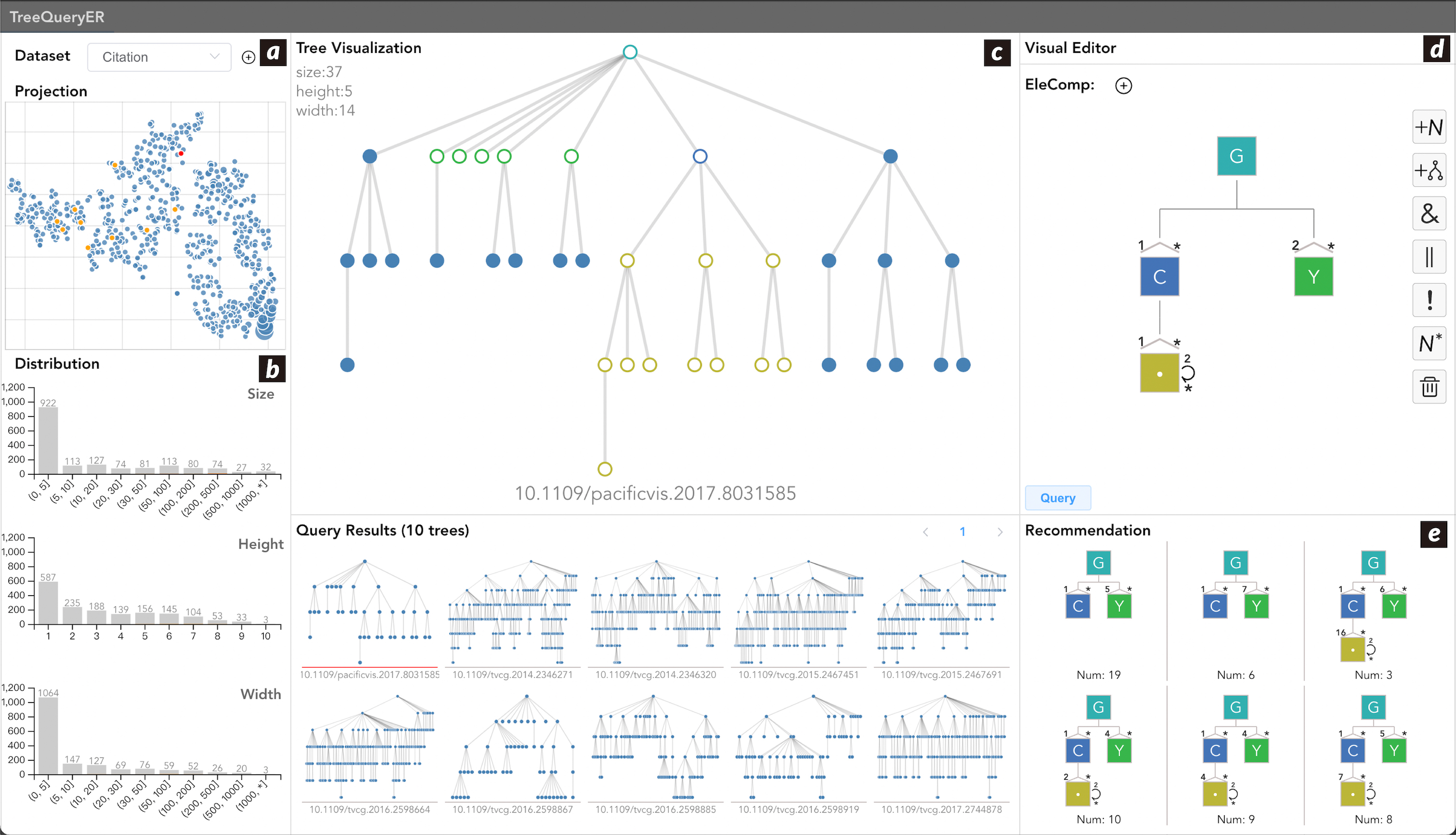}
 \caption{The user interface of the TreeQueryER prototype system. (a) data collection overview panel. (b) data distribution panel. (c) tree visualization panel. (d) visual editor panel. (e) expression recommendation panel.}
 \label{fig:system-interface}
\end{figure*}

To address the above limitations, we construct a semantic overview based on the topological structures of hierarchical data using a graph embedding method. 
Our goal is to learn a low-dimensional representation that captures the structural information of the graph, ensuring that graphs with similar structures are adjacent in a two-dimensional space. 
To achieve this goal, we first merge hierarchical data with identical topological structures and then apply a graph contrastive learning method (GraphCL)~\cite{you2020graph} to map the structural information of graphs into high-dimensional vectors. 
GraphCL employs a contrastive loss function to maximize the consistency between positive pairs in comparison to negative pairs.
Fig.~\ref{fig:pipeline}(c) illustrates the detailed architecture of the framework. 
GraphCL augments each hierarchy in the dataset by randomly dropping a node and its subtree to construct positive pairs. 
To ensure that the hierarchical structure is not significantly changed, the height of the dropped node should be less than or equal to two. 
Considering the node matching process of HiRegEx is from top to bottom, we design node attributes as \textit{depth} because the changes in the node depth can significantly influence the topological structure.  
Subsequently, we apply the t-SNE dimensionality reduction algorithm~\cite{van2008visualizing} to project the high-dimensional latent vectors of hierarchical data into a two-dimensional space for an overview. 
From the overview, analysts can understand the similarities between any pairs of hierarchical data and identify the patterns/anomalies.

\section{TreeQueryER Prototype System}

We have designed and implemented the TreeQueryER prototype system to facilitate the exploratory visual analysis for multivariate hierarchical data based on the HiRegEx grammar. 

\subsection{Design Consideration}

\textbf{DC1: Reducing the cognitive burden for constructing query expressions based on the HiRegEx specification.}
Constructing HiRegEx expressions in a textual format has a steep learning curve. 
Specifically, users need to memorize the operators and parameters in the HiRegEx specification. 
The prototype system should enable users to analyze multivariate hierarchical data effectively and efficiently. 
However, manually writing textual query expressions in textual format contradicts this goal. 
More specifically, the manual construction process is time-consuming and query expressions are not intuitive. 
Inspired by various visual query studies for graph data~\cite{troidl2023vimo}, event sequence~\cite{cappers2018eventpad}, movement sequence~\cite{krueger2017vespa}, and temporal pattern~\cite{krause2015supporting}, which support user interaction for constructing visual query patterns intuitively, TreeQueryER also aims to allow users to construct query expressions through direct manipulation and display them in a visual format.

\textbf{DC2: Enabling users to achieve the comprehensive analysis of multivariate hierarchical data.} 
Different users conduct data analysis with varying intentions: some have explicit goals and tasks, others may have no specific objectives, and some possess vague goals with a few initial tasks~\cite{battle2019characterizing}, leading to different data analysis requirements.
During the analysis process, users' interests may refine or evolve as they observe and discover new insights, ultimately seeking the desired information~\cite{idreos2015overview, keim2001visual, wongsuphasawat2015voyager}. 
This process is a key aspect of exploratory analysis. 
Therefore, it is essential to implement various exploration modes, such as top-down, bottom-up, and context-creation (introduced in Sec.~\ref{sec:query-based-exploration-framework}), to support diverse analysis needs~\cite{battle2019characterizing}. 
These modes should be integrated into the system to realize comprehensive data exploration.

\subsection{User Interface and Interaction}
The user interface of the TreeQueryER system is shown in Fig.~\ref{fig:system-interface}, and it includes a visual editor panel, a data overview panel, an expression recommendation panel, and a tree visualization panel.

\begin{figure*}[tb]
 \centering 
 \includegraphics[width=\textwidth]{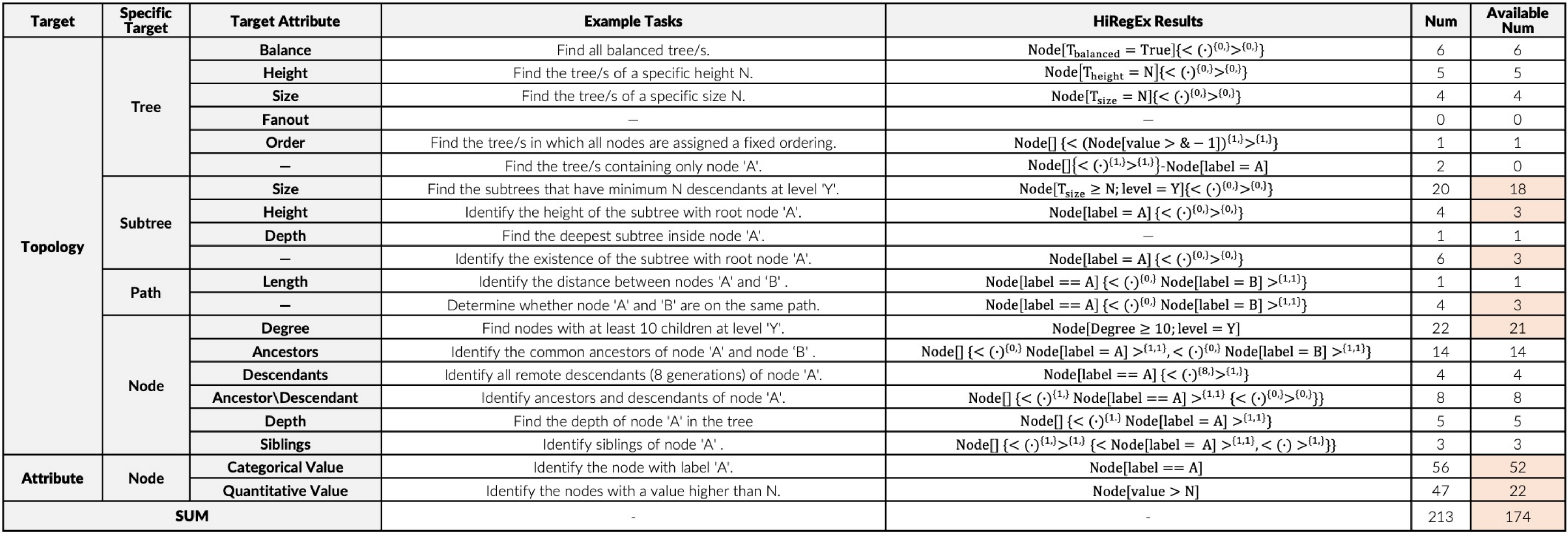}
 \caption{The multi-level task topology of the e-MLTT framework, which consists of three levels, target (high-level), specific target (mid-level), and target attribute (low-level). Each category provides a representative task and the corresponding query expression using the HiRegEx grammar. The cells with a yellow background color indicate task categories that HiRegEx cannot fully achieve.}
 \label{fig:task-evaluation}
\end{figure*}

The \textbf{data collection overview panel} (see Fig.~\ref{fig:system-interface}(a)) is tailored to fulfill the requirements of the context-creation exploration mode. This panel demonstrates the distributions of multivariate hierarchical data through a scatter plot. 
Each node in the scatter plot signifies a sub-collection with the same topological structure. 
The scatter plot visualization encodes the amount of data in the collection into node size. 
Distances between nodes indicate similarities in hierarchical data from the topological perspective. 
The scatter plot highlights query results, providing context for visual query results and aiding users in comprehending distributions across the entire data collection (\textbf{DC2}). 
The TreeQueryER system also enables data filtering by diverse attributes (\textit{e.g.}, size, height, and width) in the distribution panel (see Fig.~\ref{fig:system-interface}(b)). 

The \textbf{tree visualization panel} displays the results of visualizing multivariate hierarchical data that match the query expression. 
We employ the same color to associate nodes with their matched elements in query expressions.
Users can click on each node within the tree visualization to inspect detailed node attributes.
At the bottom of the tree visualization panel (see Fig.~\ref{fig:system-interface}(c)), the TreeQueryER system furnishes users with thumbnails of the tree visualizations, encompassing the entire collection matched with the expression. 

The \textbf{visual editor panel} (see Fig.~\ref{fig:system-interface}(d)) facilitates the top-down exploration mode (\textbf{DC2}). It offers users an interface for constructing HiRegEx expressions through direct manipulations. The design of this panel adheres to the HiRegEx specification outlined in Sec.~\ref{sec:grammar-specification}. 
The node corresponds to a rectangle, allowing users to define constraints for multiple attributes. 
The path corresponds to multiple sequentially connected rectangles, and users can specify the repetition number of nodes. 
Similarly, the branch consists of multiple paths, and users can specify a repetition number for each path. 
Components can be dragged and connected to construct a query expression. 
Users can connect the components to denote parent-child relationships. 
The visual representation of the expression in the visual editor panel employs the node-link tree visualization, enhancing consistency with users' cognitive understanding of targets and aiding in identifying matching relationships between components and query results ((\textbf{DC1})).

The \textbf{expression recommendation panel} (see Fig.\ref{fig:system-interface}(e)) is designed to support bottom-up, data-driven inquiries (\textbf{DC2}). 
The bottom-up mode is part of a browsing-oriented process where users lack a clear target or a specific expression for the target, needing to determine the expression based on the hierarchical data collection. 
The query expression constructed by users in the visual editor panel comprises multiple parameters that are challenging to determine.
For instance, users seeking important papers may set a criterion that their children in the citation tree consist of many highly-cited papers, but determining the threshold is challenging. 
In such cases, the query expressions constructed by users represent a rough direction rather than a determined pattern for the targets. 
After users provide an initial query statement, TreeQueryER displays relevant expressions and the quantity of matching hierarchical data in the entire dataset, thereby accelerating the process of obtaining the desired expression and query results ((\textbf{DC1})).
Users can further select the recommended query expression to visualize their query results in the tree visualization panel (see Fig.\ref{fig:system-interface}(c)).

In summary, the visual editor panel, expression recommendation panel, and data collection overview panel can support the top-down, bottom-up, and context-creation modes, respectively. The tree visualization panel allows users to understand the query results to refine their query expressions. All the above panels can support users' analysis for multivariate hierarchical data.

\section{EVALUATION}

We validate the above techniques from two aspects. First, we demonstrate the expressiveness of HiRegEx based on the e-MLTT framework. Second, we validate the utility of the TreeQueryER system through a use case on the citation tree dataset in the visualization field. 

\subsection{Performance Evaluation}
\label{sec:grammar-expressiveness}
\textbf{Expressiveness of the HiRegEx grammar.}
We validated the expressiveness of the HiRegEx grammar based on the e-MLTT framework. 
More specifically, we utilized the HiRegEx grammar to specify the targets in the 213 tasks underlying the e-MLTT framework. 
We carefully recorded the number of tasks that could be supported by the HiRegEx grammar, with the results detailed in Fig.~\ref{fig:task-evaluation}. Out of the 213 tasks, HiRegEx can support 174 of them. For each category, we selected a representative task and demonstrated the application of the HiRegEx grammar to define its target. 
Of the 39 tasks that are not supported, 7 of them involve no query operations. An example of such tasks is ``comparison of different subtrees''. The remaining 32 tasks primarily concern computing ``extreme'' values, such as ``find a node having the maximum attribute value of the second layer''. 
HiRegEx does not consider aggregation operations for query results. 
This limitation affects its effectiveness in these tasks related to extreme values. 
However, this shortfall is effectively mitigated by TreeQueryER, which tools are provided for interactively filtering extreme values within the results set. 
Moreover, the experiment results revealed that HiRegEx proficiently supports tasks related to topology, signifying its capability to represent the structural attributes of hierarchical data accurately.

\subsection{Case Study}
We validate the utility of the TreeQueryER prototype system by collaborating with two expert users.
This section presents a use case on the citation tree dataset. 
This use case demonstrates that TreeQueryER can help users achieve a comprehensive exploratory visual analysis for the citation tree dataset and get insights into the development and intersection of research topics.

\textbf{Dataset.} To assemble a thorough and representative dataset for our study, we crawled a total of 1644 research papers in the field of visualization, covering the period from 2014 to 2020. 
This dataset encompasses a diverse range of publications from IEEE VIS/TVCG, EuroVis, and PacificVis. 
The attribute data of each paper include title, authors, publication date, affiliation, country, and citation list/count. 
We then processed the data collection to construct a citation tree for each paper. 
In a tree, a paper ``\textit{b}'' that cited another paper ``\textit{a}'' is a child node of ``\textit{a}''. 
This structure enabled us to visualize and analyze the intricate network of citations within the field of visualization.

\begin{figure*}[tb]
 \centering 
 \includegraphics[width=0.97\textwidth]{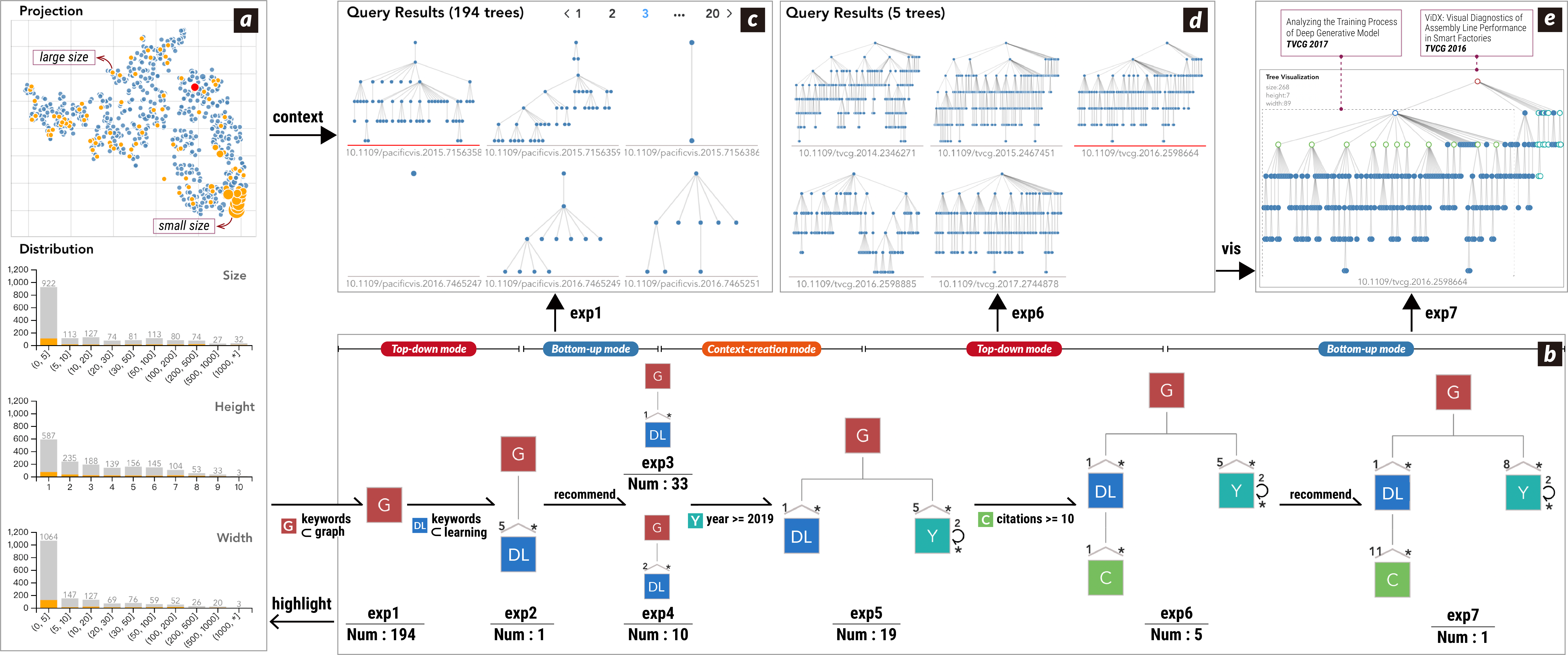}
 \caption{
The use case on the citation tree dataset. (a) The data collection overview panel offers the exploration context and highlights user query results. 
(b) The process of constructing query expressions comprises seven statements indicating the respective number of query results. 
The links between statements reveal the attributes and corresponding constraints. 
In particular, \textit{exp4} and \textit{exp7} are suggested by the recommendation algorithm; 
(c) The query results of the \textit{exp1} statement; 
(d) The query results of the \textit{exp6} statement; 
(e) The query results of the \textit{exp7} statement.
 }
 \label{fig:citation-case}
\end{figure*}

\textbf{Expert Users and Tasks.} We invited two experienced visualization researchers, referred to as E1 and E2, to evaluate the TreeQueryER system. 
Both experts have over five years of research experience in the field. 
Typically, researchers build citation graphs using tools such as Connected Papers~\cite{connectedpaper} to explore papers' relationships. 
This involves manually navigating through citation graphs to understand the evolution.
In our case study, E1 and E2 aimed to explore how research in graph visualizations, a traditional topic, intersects with deep learning using TreeQueryER. Before delving into the dataset, they received an introduction to the TreeQueryER prototype and the specifications of the HiRegEx query grammar. Then, they used TreeQueryER for exploratory visual analysis.

\textbf{Step 1: Querying papers on graph visualizations and deep learning through the top-down mode.} Initially, the experts were given a brief introduction to the citation tree dataset and various tools provided by TreeQueryER. 
Then, they decided to explore papers related to deep learning techniques and graph visualization. 
To identify papers related to graph visualization, they initialized a node in the visual editor, requiring their keywords list to include ``\textit{graph}''. The node, along with the corresponding constraints of attribute values, was labeled as \raisebox{-2pt}{\includegraphics[height=1em]{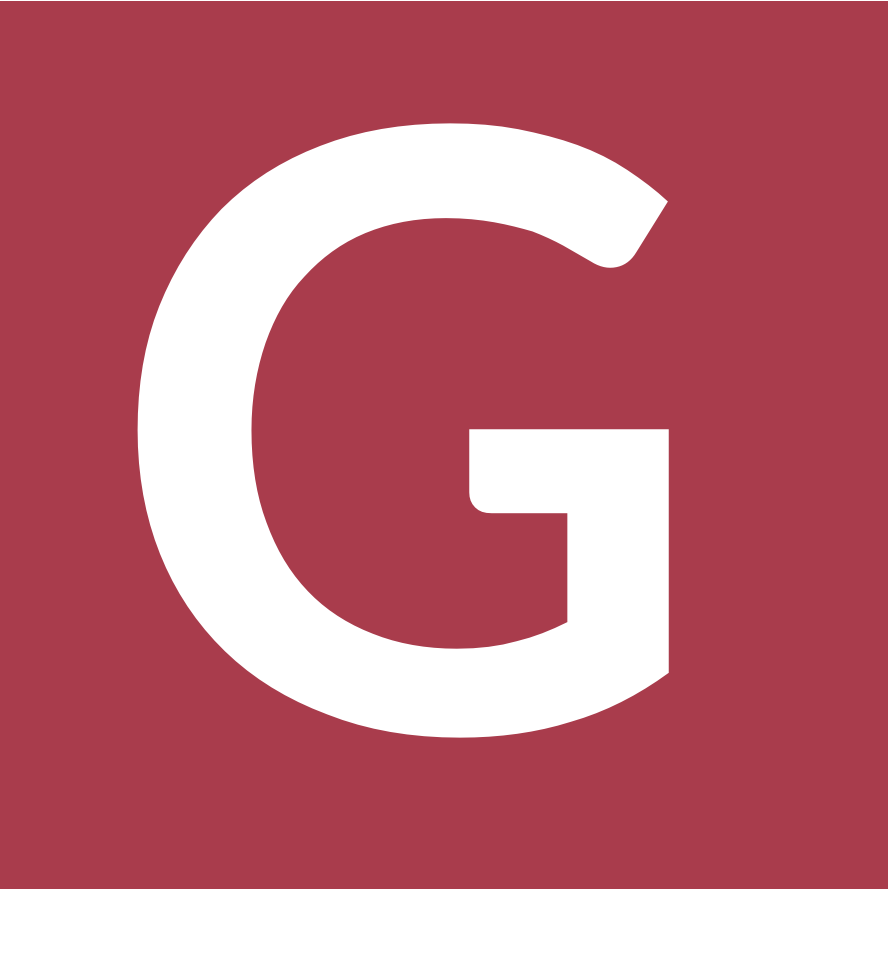}} (representing the topic ``\textbf{g}raph''), as shown in the $exp_1$ of Fig.~\ref{fig:citation-case}(b).
Employing this query expression, the experts retrieved 194 papers.

\textbf{Step 2: Querying papers cited by deep learning-related papers using the bottom-up mode.} 
Many publications may not employ deep learning techniques but still have an impact on this domain. 
These papers can provide researchers with valuable inspiration but cannot be obtained through the above query expression. 
To retrieve these papers, the experts constructed another expression to query those cited by more than five papers related to the ``deep learning'' topic. 
They added a branch with more than five repetitions in the visual editor, comprising a node labeled \raisebox{-2pt}{\includegraphics[height=1em]{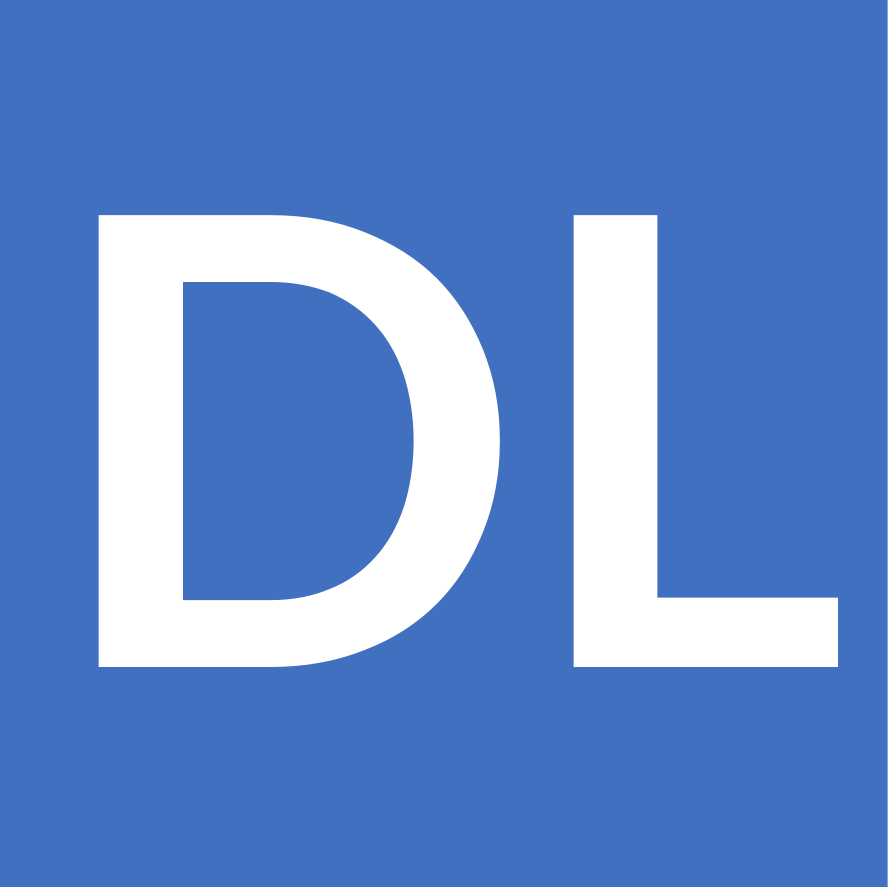}} with a constraint that the keywords included the term ``\textbf{d}eep \textbf{l}earning''.
Connecting this branch with the \textit{node} \raisebox{-2pt}{\includegraphics[height=1em]{pictures/node_graph.pdf}}, they executed the query expression. The resulting query (exp2 in Fig.~\ref{fig:citation-case}) had only one paper due to the relatively strict constraint of being cited by more than five deep learning-related papers. 
This result hindered a comprehensive understanding of the dataset. 
At this point, the bottom-up recommendations of the HiRegEx expression became pivotal.
From the expression recommendation panel, experts ascertained that the number of papers cited by one or two deep learning-related papers was 33 and 10. 
Based on these recommendations, they adjusted the parameter of branch repetitions in the query expression to enhance the diversity of results. 
Fig.~\ref{fig:citation-case}(c) presents the query results. 
The projection panel is updated accordingly, highlighting circles that match the query results in yellow, as depicted in Fig.~\ref{fig:citation-case}(a).

\textbf{Step 3: Identifying anomalies through the context-creation mode.} The researchers identified distinct clusters by examining the distributions in the projection panel. 
From the projection view, they learned that trees located in the lower right corner exhibited small sizes, in contrast to the larger trees represented by circles in the upper left corner, as shown in Fig.~\ref{fig:citation-case}(a). 
Based on this observation, they selected a cluster of interest for in-depth analysis.
Upon closer inspection, the researchers discovered that certain citation trees consisted of only a few levels, suggesting that the research topics of these papers were outdated and lacked continued exploration by researchers. 

\textbf{Step 4: Refining the query expression interactively.} Furthermore, they introduced a new branch under the node \raisebox{-2pt}{\includegraphics[height=1em]{pictures/node_graph.pdf}}, complementing the constraint that the descendants must comprise more than five papers after 2019 (denoted as \raisebox{-2pt}{\includegraphics[height=1em]{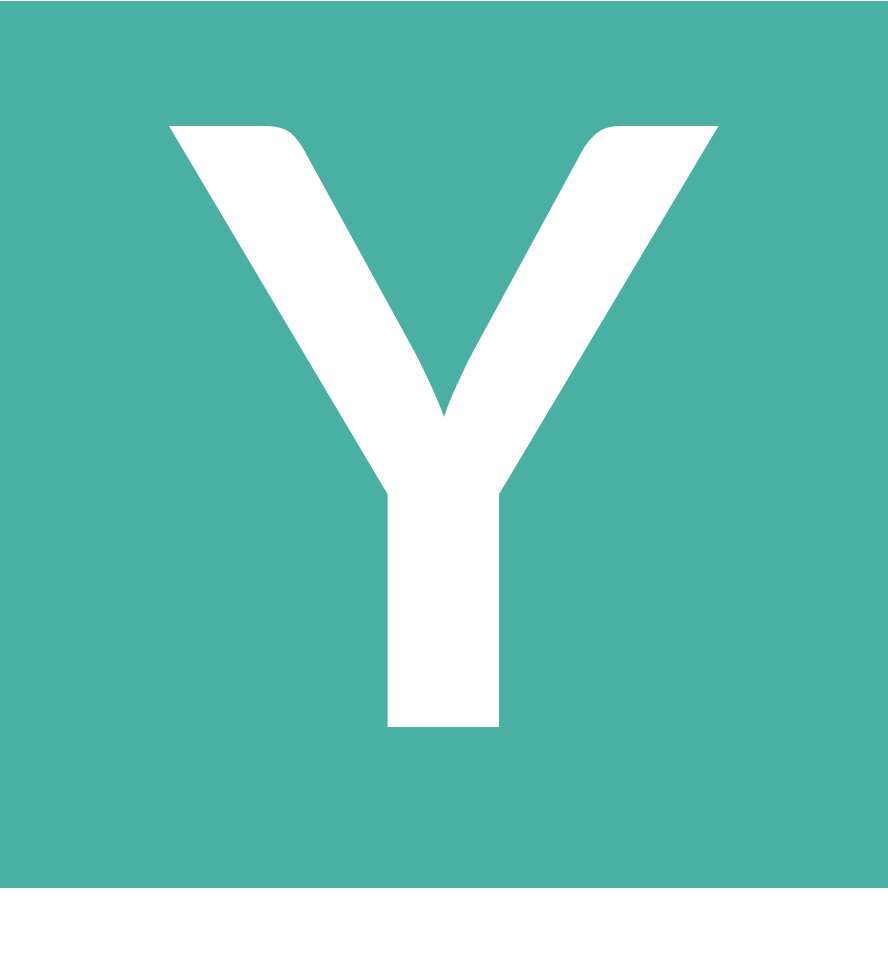}}). 
These constraints were implemented to ensure that the papers in the query results remain within an active research area.
Executing this refined query expression yielded 14 citation trees. Notably, four trees within the results lacked subsequent citations, indicating a limited impact. 
To measure a paper's impact, they considered its citations and whether highly-cited papers referenced it. 
Accordingly, they set a constraint for the ``citation'' attribute of ``node'' to be greater than ten (denoted as \raisebox{-2pt}{\includegraphics[height=1em]{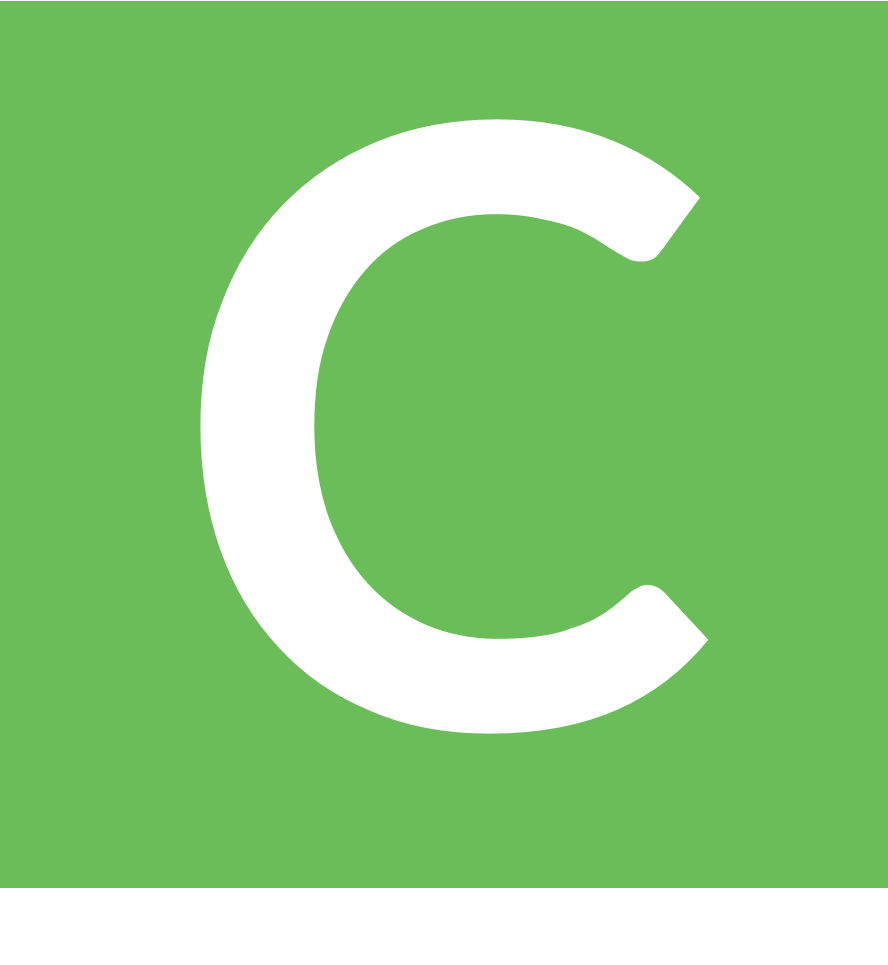}}). 
Next, they introduced a new branch to restrict papers cited by at least one highly-cited paper (exp6 in Fig.~\ref{fig:citation-case}).
Note that the parameters in the above expression are adjusted by the recommendation algorithm based on the dataset without users' manual specifications. 

From the expression recommendation panel, experts identified an expression (exp7 in Fig.~\ref{fig:citation-case}) corresponding to a specific query result (ViDX~\cite{xu2016vidx}), which emerged as the most relevant result for the given expression. This result pertained to ``graph'' and was cited by influential papers on ``deep learning''.
After retrieving the paper, they explored the projection view and found that many nearby trees contain another influential paper, \textit{``Analyzing the Training Process of Deep Generative Model''}~\cite{liu2017analyzing}. 
They learned that the outlier detection capability of visual analytics, such as the methods based on Marey’s Graph in the smart factories usage scenario, can be used to help users identify the outlier causing a failed training process.

\textbf{Expert Feedback.} 
We conducted one-on-one 30-minute interviews with the two experts mentioned above to gather their feedback on our techniques after they finished their hand-on explorations. 
We encouraged them to freely share their thoughts on our methods as well as their impressions of the overall experience.
Both experts expressed positive attitudes towards our method and agreed that the system could improve the efficiency of the paper-searching process.
More specifically, they appreciated the design of HiRegEx.
``\textit{HiRegEx allows me to flexibly define various conditions, such as citation counts and papers' relations. While each condition alone may not be complex, combining them can be quite intricate. HiRegEx provides an intuitive and user-friendly way to describe these queries (E1)}''.
HiRegEx extends the basic usage of regular expressions, making it easy for them to understand the rules and use the grammar conveniently. 
Furthermore, they appreciated the ability to construct their query expressions effectively using the visual editor panel of TreeQueryER. 
``\textit{The visual representation of HiRegEx expressions is intuitive, allowing me to quickly turn an idea into an expression. Through interactive manipulation, I can easily construct the query expressions I need (E1, E2)}''.
The visual cues enabled them to recognize the characteristics of their query statements. 
In addition, they found the visual query expression recommendation approach beneficial for refining their queries and selecting appropriate parameters. 
``\textit{The recommendation mechanism always provides me with effective suggestions for the next-step exploration (E2)}''.
The projection view offered a semantic overview, presenting the context of query results and facilitating the identification of patterns for further investigation.

\section{DISCUSSION AND FUTURE WORK}

\textbf{Comparison with Existing Query Grammars.} The techniques competing with
HiRegEx for hierarchical data query includes Tregex~\cite{levy2006tregex}, Jaql~\cite{beyer2011jaql}, and HQL~\cite{slingsby2009configuring}.
For graph data, the techniques Gremlin~\cite{rodriguez2015gremlin}, Cypher~\cite{francis2018cypher}, PGQL~\cite{van2016pgql}, and G-CORE~\cite{angles2018g}. 
These existing techniques differ in motivation, expressiveness, available tutorials, and prototype systems, which involve many confounding variables.
Hence, we did not conduct a quantitative study to evaluate the efficiency of HiRegEx grammar. 
Most existing studies predominantly focus on node attributes, overlooking tree-specific attributes such as size, height, and depth. This limitation hampers their capability to effectively describe large trees, as their query languages necessitate detailed node specifications to define a tree structure accurately.
For instance, Tregex~\cite{levy2006tregex}, tailored for syntax trees, adeptly specifies topological structures and node attributes but cannot define constraints from the perspective of tree compositions. 
In contrast, Jaql~\cite{beyer2011jaql} and HQL~\cite{slingsby2009configuring}, designed for semi-structured hierarchical database queries, exhibit predetermined topology in their query targets, restricting users from defining query patterns flexibly.
Gremlin~\cite{rodriguez2015gremlin}, Cypher~\cite{francis2018cypher}, PGQL~\cite{van2016pgql}, and G-CORE~\cite{angles2018g} are graph query languages. However, these techniques only partially consider tree-specific characteristics. PGQL~\cite{van2016pgql} and G-CORE~\cite{angles2018g} can support the specification of the sibling relationships of hierarchical data. However, they do not support queries for a large tree because they need to specify the query targets in a fine-grained manner.

\textbf{Expressiveness of the HiRegEx Grammar.} The e-MLTT framework, dedicated to tree visualizations, encompasses 213 analysis tasks. HiRegEx can support 174 of them, as shown in Fig.~\ref{fig:task-evaluation}. We have categorized the unsupported analysis tasks into two distinct groups. Tasks falling within the first category necessitate user interactions, aggregation, and computation after querying. 
These tasks involve actions related to finding the extreme value or aggregating data, exemplified by queries like ``What is the maximum depth of the hierarchy?'' or ``How many files are there in the directory?''. Another common example entails determining the least common ancestor of two nodes.
In contrast, tasks in the second category are incompatible with query tasks. 
An illustrative example includes tasks centered around assessing the balance of trees or subtrees in hierarchical data.

\textbf{Further Improvements of the Construction Efficiency.} Considering various operators and constraints for the node attributes, the query expression of HiRegEx cannot be completely expressed or constructed in the textual format like the traditional regular expressions. 
Therefore, we plan to develop a library to integrate the HiRegEx expression with popular programming languages, like Python, to improve the utility of the HiRegEx in textual format. 
Another future work is to improve user construction efficiency for query expressions by natural language processing techniques.
We will explore natural language processing techniques to generate the corresponding query expression and expression description based on Large language models (LLMs)~\cite{li2024llmevaluation} in the future.
To improve readability, we also explore providing a short natural language description of the expressions constructed by users.

\section{Conclusion}
We present HiRegEx, a declarative grammar designed for querying multivariate hierarchical data. 
HiRegEx borrows the operators from the classical regular expressions and further extends their expressiveness according to the characteristics of multivariate hierarchical data. 
Based on HiRegEx, we developed a query-based exploratory framework, which consists of \textit{top-down} pattern specification, \textit{bottom-up} data-driven inquiry, and \textit{context-creation} data overview.
We implemented a prototype system, TreeQueryER, to integrate our exploratory framework. 
We validate the expressiveness of HiRegEx based on the e-MLTT framework.
We also demonstrate the effectiveness and utility of the exploratory framework and TreeQueryER prototype system through a case study involving a citation tree dataset.

\acknowledgments{%
This work is supported by National Key R\&D Program of China (2021YFB3301500), NSFC (62302038, U2268205), Young Elite Scientists Sponsorship Program by CAST (2023QNRC001), and Tencent Rhino-Bird Focused Research Program.
}

\bibliographystyle{abbrv-doi-hyperref}

% \bibliography{VQTree}

\appendix %

\end{document}